\documentclass[lettersize,journal]{IEEEtran}

\IEEEoverridecommandlockouts

\usepackage{forest,array}
\usepackage{tikz}
\usepackage{hyperref}

\hypersetup{colorlinks=true,	linkcolor=blue,	citecolor=blue,	}
\usepackage{cite}
\usepackage{amsmath,amssymb,amsfonts}
\usepackage{algorithmic}
\usepackage{graphicx}
\usepackage{textcomp}
\usepackage{xcolor}
\usepackage[none]{hyphenat}
\usepackage{enumitem}

\setlist{leftmargin=4mm}
\begin{document}

\title{Cybersecurity Challenge Analysis of Work-from-Anywhere (WFA) and Recommendations  guided by a User Study
  } 
\author{Mohammed Mahyoub, Ashraf Matrawy, Kamal Isleem, Olakunle Ibitoye \\School of Information Technology, Carleton University, Ottawa, Canada
    \thanks{$^{\S}$The corresponding author, Email: mohammed.mahyoub@carleton.ca.

}
}

\maketitle
\thispagestyle{plain}
\pagestyle{plain}
\begin{abstract}
Many organizations were forced to quickly transition to the work from anywhere (WFA) model as a necessity to continue with their operations and remain in business despite the restrictions imposed during the COVID-19 pandemic. Many decisions were made in a rush, and cybersecurity decency tools were not in place to support this transition. In this paper, we first attempt to uncover some challenges and implications related to the cybersecurity of the WFA model. 
Secondly, we conducted an online user study to investigate the readiness and cybersecurity awareness of employers and their employees who shifted to work remotely from anywhere. The user study questionnaire addressed different resilience perspectives of individuals and organizations. The collected data includes $45$ responses from remotely working employees of different organizational types: universities, government, private, and non-profit organizations. 
Despite the importance of security training and guidelines, it was surprising that many participants had not received them. A robust communication strategy is necessary to ensure that employees are informed and updated on security incidents that the organization encounters. Additionally, there is an increased need to pay attention to the security-related attributes of employees, such as their behavior, awareness, and compliance. Finally, we outlined best practice recommendations and mitigation tips guided by the study results to help individuals and organizations resist cybercrime and fraud and mitigate WFA-related cybersecurity risks.


\end{abstract}


\section{Introduction}
The work-from-anywhere (WFA) model has become increasingly popular in recent years, allowing employees to work remotely and have more flexible schedules. A Gartner survey \cite{b1} reveals that more than $74\%$ of organizations will support a certain population of their workforce to continue working from home or anywhere. 
However, this model introduces new risks and challenges that transform the cybersecurity landscape. Attackers may exploit the vulnerabilities that arise from employee behavior and employer policies \cite{b0}.  Remote employees may connect to unsecured public WiFi networks, use shadow IT \cite{b2}, and personal devices to access corporate resources or store sensitive data on unsecured devices, increasing the risk of data breaches and cyber-attacks. Phishing attacks, malware infections, and insider threats can also be more challenging to detect and prevent when employees work remotely. Additionally, compliance with regulatory requirements can be challenging when employees work remotely. By understanding these challenges and implementing appropriate cybersecurity measures, organizations can ensure that the WFA model is a safe and secure environment for their employees and their data. However, despite these increasing risks, there is a well-documented discrepancy between individuals' privacy concerns and their actual behaviour, often referred to as the privacy paradox as reviewed by Gerber et al. \cite{b4a}. This can be explained by several factors, including behavioural biases (such as the intention-behaviour gap), a lack of awareness of privacy risks, and trade-offs between privacy and other perceived benefits, such as convenience. In the context of WFA, this paradox is particularly relevant, as remote employees may express concerns about cybersecurity but still engage in risky behaviours, such as using public networks without using VPN due to convenience or lack of perceived immediate threat.
In light of the above discussion, the contribution of this paper is threefold:
\begin{itemize}
\item We analyzed the cybersecurity challenges and implications of the WFA model and categorized them based on the model part they belong to. This can provide organizations and individuals with a better understanding of the potential risks and vulnerabilities associated with the WFA model and improve the appropriate incident response management accordingly. 
\item  We conducted an online user study to evaluate the cybersecurity awareness and readiness of organizations and their employees when they switch from on-site to remote work. The study aimed to provide valuable insights that can help organizations and individuals take proactive steps to protect themselves and their employees from cybersecurity threats by identifying knowledge gaps, assessing employee behavior, evaluating security policies, identifying areas for improvement, and promoting best practices.

\item  We outlined best practice recommendations and mitigation tips guided by study results. They aim to aid individuals in resisting cybercrime and fraud, help organizations reduce cybersecurity risks, strengthen their security posture, and improve compliance with relevant regulations.
\end{itemize}
The remainder of this paper is structured as follows. Section \ref{sec:relted_work} reviews the related work. Section \ref{sec:wfa_overview} presents the  WFA model components and its cybersecurity challenges and implications. Section \ref{sec:methodology}  presents the design methodology, demographic, and validity analysis of the user study. The data collected are analyzed in Section \ref{sec:results_and_analysis}. In Section \ref{sec:limitations}, the key findings and limitations of this study are given. Section \ref{sec:recommenations} presents best practice recommendations for effective cyberattack risk reduction. Finally, Section \ref{sec:conclusion} concludes the paper.

\section{Related Work}
\label{sec:relted_work}
Although the cybersecurity challenges related to the WFA model have been discussed online in blogs and news articles in different domains, we tried to categorize these challenges in more structured analyses in this paper.
Few user studies have been conducted and published that assessed remote workers' and organizations' readiness and cybersecurity awareness. This section reviews the most notable related works to our study.

Nyarko \textit{et al.} in \cite{Nyarko2023} surveyed to assess the compliance of remote workers with cybersecurity regulations established by their respective organizations. The survey included participants from the UK, where the majority (67.8\%) of respondents reside. The researchers concluded that more in-depth research is necessary to understand the evolving cybersecurity landscape and the level of employee engagement with cybersecurity policies, procedures, and training programs.
\emph{Our investigation focused on Canadian residents as participants and explored several aspects, including employer security precautions, availability of security-enhancing resources/tools, provision of security awareness training, and overall security awareness level of employees.}

 Mannebäck \textit{et al.} in \cite{16}  ethnographically studied the effect of the change in the workplace during the COVID-19 pandemic. They mainly targeted IT people in charge of securing the communication and network for remote workers. The study was carried out in partnership with a specific team inside one of Sweden's largest counties that specializes in IT. They collected data by analyzing documents, diaries, and transcripts of focus group sessions. Given that the research was confined to a single department within a particular organization and did not encounter any apparent cyberattacks during the pandemic, the study findings cannot be generalized. \emph{Our study, however, targets different IT knowledge levels of remotely working employees regardless of the department or the organizational type that they are working for}.
 
 Georgiadou \textit{et al.} in \cite{15} intended to assess the readiness of firms from $13$ European nations and various fields of business regarding cybersecurity culture when the COVID-19 pandemic required teleworking. They focused on workers' feelings, opinions, viewpoints, and uniqueness, as employees are the first line of defense for any system. 
To understand the issue comprehensively, our study considers additional factors such as organizational awareness, behavior, and availability of security-enhancing resources and tools. Additionally, our study focuses primarily on employees in Canada who work remotely.

The Strategic Council of the Canadian Internet Registration Authority (CIRA) published the user study findings in \cite{18}. It collected $510$ online responses from cybersecurity decision-makers across Canada. The study was conducted in July and August 2021 with a focus on identifying what organizations have done to mitigate remote work-related cybersecurity challenges during the COVID-19 pandemic. Their study targeted decision-makers to evaluate their organization's performance. \emph{Our study, on the other hand, also asked about the different rules of employees, and questions to find more detailed information about issues that could lead to cybersecurity vulnerabilities, such as training, setting up the workplace and environment, providing equipment/resources to employees, and the awareness and reaction of the employee in case cyberthreats occur}. Therefore, to summarize, the main aim of their study was to identify to what extent organizations are prepared to shift to work from home from a cybersecurity perspective. In addition to the above purpose, our study evaluates employees' cybersecurity awareness and their behavior when threats are detected.

Cybersecurity Insiders \cite{18}  conducted a survey in May $2020$. It polled $413$ IT security decision-makers, practitioners, and companies of varying sizes in multiple industries. The study tried to evaluate to what extent organizations are prepared to shift from an on-premises to a remote workforce. \emph{Our study spans all organization employees and asks for more details to determine their readiness for transformation to the WFA scheme. Furthermore, specific questions in our study try to challenge employees to evaluate their cybersecurity awareness while working remotely}.  

Lallie \textit{et al.} \cite{b26} investigated the correlation between global events during the COVID-19 pandemic and cyberattacks. They reported a small correlation between the authority-announced policies for the pandemic and related cyberattack campaigns that use these events as a hook, increasing the chance of success. According to their timeline research, many cyberattacks start with a phishing effort that instructs victims to download a file or visit a URL. The file or URL functions as a delivery vehicle for malware, which, once installed, becomes a vehicle for financial theft. The investigation also revealed that the phishing effort uses media and government statements to increase its chances of success.
Furthermore, governments, the media, and other institutions should be aware that announcements and the publication of stories will likely lead to the launch of related cyberattack activities that take advantage of these events. The events should be accompanied by a message or disclaimer detailing how the announcement's contents will be disseminated. Although the work referenced in \cite{b26} is not directly related to our work in this paper, it sheds light on the fact that the WFA model has been targeted by numerous cyberattacks during the pandemic. Therefore, it is important for organizations to be aware of these threats and take appropriate measures to mitigate them.

\section{WFA Model, Challenges, and Implications}
\label{sec:wfa_overview}
This section presents the WFA model, common cybersecurity challenges, and implications related to the WFA model.

\subsection{WFA Model}
The WFA model  presented in this paper  has four main components; 
employees, devices, software, connectivity networks, and an organization (i.e., an employer). Each component would cause various challenges and, therefore, should have different roles to prevent cyberthreats. In the following, these components are defined:
\begin{itemize}
    \item \textbf{Employees:} Over 90\% of security breaches are caused by human mistakes \cite{b6}.
    Employees have a varying level of responsibility to protect the organization against cyberattacks. A misconception that exists sometimes leads some employees to think that cybersecurity tasks are the job of IT experts and technology professionals solely. Although employees' cybersecurity responsibilities may not have been fully clear before the COVID-19 pandemic, now organizations are more aware than ever of them \cite{b9}.
    Indeed, the way that employees conduct their daily IT activities while working remotely is crucial to the security of the entire organization's system. 
    
    \item \textbf{Devices and Software:} With WFA, a wide range of devices could be used for remote work, such as workstations, laptops, smartphones, and tablets. Different software and applications are installed to enable employees to complete their daily activities and duties properly. Examples are virtual private network (VPN) software, videoconferencing tools, group chat software, etc. 
    \item \textbf{Connectivity Network:} It is the path through which employees access their organizational networks or data centers. In order to protect the operation of the WFA model, this path should be secured, and data has to pass through trusted devices and networks starting from the home or the public router and ending at the organization's gateway. In practice, providing a guarantee for complete end-to-end security is almost impossible.
    \item \textbf{Organization:} Protecting the WFA model is in the best interest of any organization that adopts this model, as operational or security breaches to this model could be very costly for the organization. Therefore,  organizations have to ensure, among other things, that their employees are well-trained to protect their devices and have a secure connection. This is in addition to the typical responsibility of securing the organization's network, devices, and applications. 
\end{itemize}

\subsection{WFA-related Cybersecurity Challenges}

\begin{figure}[!ht]
    \centering
    \scalebox{0.6}{
        \begin{forest}
            for tree={
                if level=0{
                    align=center,
                }{
                    align={@{}lp{45mm}@{}},
                },
                grow=east,
                draw,
                font=\sffamily\bfseries,
                edge path={
                    \noexpand\path [draw, \forestoption{edge}] (!u.parent anchor) -- +(5mm,0) |- (.child anchor)\forestoption{edge label};
                },
                parent anchor=east,
                child anchor=west,
                l sep=10mm,
                tier/.wrap pgfmath arg={tier #1}{level()},
                edge={ultra thick, rounded corners=2pt},
                fill=white,
                rounded corners=2pt,
            }
            [{WFA Challenges}, rotate=90
                [{\hyperref[sub:orgChallenges]{Organization}},
                    [Third-party IT providers vulnerabilities]
                    [Identity vulnerabilities]
                    [Restricted IT support]
                ]
                [{\hyperref[sub:netChallenges]{Network}},
                    [Insider threats of untrusted networks]
                ]
                [{\hyperref[sub:deviceChallenges]{Device \& Software}},
                    [Shared devices]
                    [Video conferences tools]
                    [File-sharing risks]
                    [Shadow IT]
                    [Physical security]
                ]
                [{\hyperref[sub:empleeChallenges]{Employee}},
                    [Disruption]
                    [Off-boarding]
                ]
            ]
        \end{forest}
    }
    \caption{WFA Challenges}
    \label{fig:WFH-Challenges}
    \vspace{-.2cm}
\end{figure}
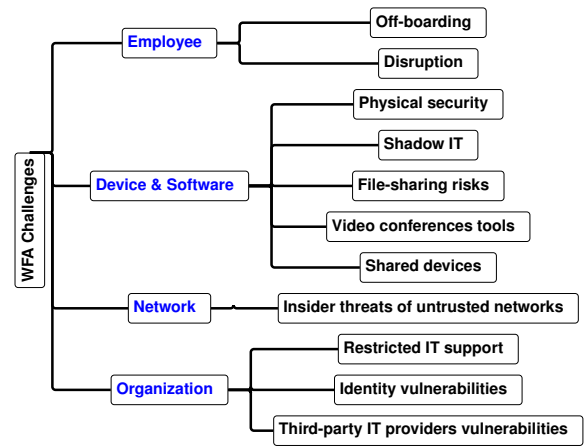

Organizations are facing new cybersecurity challenges related to WFA practices that were not previously considered on such a large scale. These challenges are forcing organizations to review their cybersecurity policies and measures. Figure \ref{fig:WFH-Challenges} outlines some of these new challenges, which can be broadly grouped into four categories based on the part of the WFA model to which they are related. These challenges could arise from employees themselves, the tools they use (such as devices and software), the communication network, or the organization.
In the following, these challenges are explained in some detail. 

\subsubsection{Employee-related Challenges}
\label{sub:empleeChallenges}
\begin{itemize}

    \item \textbf{Off-boarding} employees after the termination of their contracts present significant challenges for organizations. In the event of redundancy or termination, it is important to ensure that the employee no longer has access to or control over any information belonging to the organization. Since employees have their home offices, there is a greater risk that they may store confidential documents locally, which poses cybersecurity risks to the organization. It is worth noting that while this challenge is not unique to WFA, the risks associated with off-boarding increase with remote work practices.
    \item \textbf{Disruption}: 
    There is a lot of disruption caused by the environment surrounding remote workers.    
    The disruption may come from kids at home or from crowded people in public places. As a result, employees may lose some of their attention and make mistakes such as, for example,  sending sensitive information mistakenly to someone who shouldn't see that information or opening a phishing email \cite{b34}. Therefore, this disruption may increase the organization's security vulnerabilities.  $47$ \% of people who work from home fall victim to phishing scams, which may be due in part to inattention \cite{b35}.
\end{itemize}
	
\subsubsection{Device- or Software-related Challenges}
\label{sub:deviceChallenges}
\begin{itemize}
    \item \textbf{Physical security of devices}: Employees working outside the office have their work computers exposed to people outside the organization. The risk of an information breach is higher when an adversary has physical access to a computing device. The ability to access the device increases even as employees work from anywhere.
    \item \textbf{Shadow IT}: There has been a surge in shadow IT since the start of the COVID-19 pandemic. Shadow IT refers to the use of devices and software without the knowledge and oversight of the organization’s IT department. Many employees turn to unauthorized third-party software to complete their jobs remotely. A report published by Awake Security \cite{b2a} revealed that unauthorized remote access tools increased by $75\%$ in the first quarter of $2020$. The surge in shadow IT goes beyond unauthorized software installation and includes using unauthorized personal devices, cloud services, and Internet of Things (IoT) platforms. Shadow IT increases cybersecurity risks in an organization, such as data exfiltration and leaks, as well as non-compliance with laws and regulations \cite{b2a}.

    \item \textbf{File-sharing risks}: With several employees working from several locations, the problem of sharing files is more evident. File sharing is one of the biggest security concerns for organizations, as indicated by the Remote Working Cybersecurity Report \cite{13}. Those untrusted file-sharing platforms would expose the organization's information, leading to loss or stolen data.
    \item \textbf{Privacy violation associated with video conferencing tools}: An example of this risk is the flaw discovered in the Zoom application \cite{12}. It enables the attacker to record the Zoom sessions without notifying the participants.
    \item \textbf{Shared devices}: There are risks associated with using work devices for personal tasks or laptops for work activities. This blurred line between professional and personal lives would create new vulnerabilities. According to a report published by HP inc. \cite{b7}, $70\%$ of office workers surveyed admit to using their work devices for personal tasks, while $69\%$ use personal laptops or printers for work activities. Hackers exploit these shifting patterns to ease their phishing campaigns.
\end{itemize}
	
\subsubsection{Network-related Challenges}
\label{sub:netChallenges}
\begin{itemize}
    \item \textbf{Eroded network security perimeter:} Bring Your Device (BYOD) \cite{b38}, cloud services usage, file sharing platforms, and unsecured networks have significantly eroded the traditional network security perimeter of organizations \cite{b39}. These technologies introduce new security risks, including device and data security vulnerabilities, access control issues, compliance risks, and attempts at unauthorized access. Employees working from untrusted networks, such as coffeehouses and public WiFi networks, are vulnerable to insider threats that can compromise network security.
    
\end{itemize}
    
\subsubsection{Organization-related Challenges}   
\label{sub:orgChallenges}
\begin{itemize}

    \item \textbf{Restricted virtual IT help-desk support:} The move towards WFA has led to a rise in virtual IT help-desk support. However, the lack of in-person interaction with human IT staff can introduce cybersecurity challenges. According to a Microsoft research survey \cite{b40}, more than 40\%  of the respondents had no support staff at their location. Furthermore, more than $45\%$ of the respondents admitted to seeking the help of friends or family members to resolve computer problems, which can increase vulnerabilities \cite{14}.
    

    \item \textbf{Identity vulnerabilities:} Defining an organizational security perimeter is more challenging than ever due to the increasing use of distributed resources by employees across different devices, applications, and networks. As a result, identity vulnerabilities have become a major concern. According to ESG research \cite{10}, user identity exploitation is now one of the most common types of attacks. To combat modern threats, organizations must adapt their security defenses to be more robust and contextually relevant.

    \item \textbf{Third-party IT providers vulnerabilities:} Third-party IT providers can pose significant cybersecurity challenges \cite{10a}. Vulnerabilities in these providers can lead to unauthorized access to sensitive data, supply chain attacks, exploitation of insecure infrastructure, lack of security updates, limited control and visibility, and shared risks among multiple organizations. These risks highlight the need for thorough assessments, heightened security measures, and proactive monitoring to mitigate the potential impact on the security and integrity of work-from-anywhere setups.
    
\end{itemize}

\subsection{WFA-related Cybersecurity Implications}
As a result of the new challenges, we discuss  some of the trends that could be directly attributed to WFA, including: 
\begin{itemize}
    \item \textbf{Phishing attacks increased:} As reported in a recent survey \cite{b3}, $53\%$ of participants confirmed an increase in phishing attacks since the COVID-19 pandemic started. Another recent study \cite{b4} reported that since the end of February $2020$, the increase in phishing emails was $600\%$. Furthermore, Google blocked over $18$ million COVID-19-related phishing emails and malware every day \cite{b19}.
    \item \textbf{Data breaches increased:} This refers to incidents in which an unauthorized entity steals data from the owner. Credit card numbers, client data, trade secrets, and national security information are examples of sensitive, proprietary, or confidential information that could be stolen. The number of stolen confidential personal and organization credentials has doubled in $2020$ compared to $2019$ \cite{b32}. Furthermore, over $18.8$ billion in confidential records were exposed in the first half of $2021$ \cite{b32}.
    \item \textbf{Malware increased:} According to Deep Instinct \cite{b20}, the number of attempted malware attacks climbed by $358\%$  overall in $2020$, while ransomware increased by $435\%$ compared to $2019$.
    \item \textbf{Online scams increased:} The current COVID-19 epidemic has significantly impacted cybersecurity. According to the multi-national legal firm Reed Smith\cite{b21}, online scams increased by more than $400\%$  in March $2020$ compared to previous months.
    \item \textbf{DDoS attacks  increased:} According to NETSCOUT data\cite{b22}, distributed denial-of-service (DDoS) attacks increased significantly in $2020$ as a result of COVID-19 digital transformation to remote work. The ATLAS Security Engineering and Response Team (ASERT) cybersecurity team reported more $10$ million attacks in $2020$ which is $1.6$ million more than in $2019$.
\end{itemize}

\section{User-Study Design, Demographic, and Validity Analysis}
\label{sec:methodology}
\subsection{User-Study Designing and Testing}
The user study conducted in this work attempts to address the following questions in the context of WFA: \textbf{ Q1: \textit{ What are the security precautions employers are practicing?}},  \textbf{Q2: \textit{What is the level of security awareness among employees regarding security threats and best practices?}}
\textbf{ Q3: \textit{ To what extent can employee behavior affect the overall security of the organization?}}. 
To obtain answers to our research questions, we conducted an anonymous online user study approved by  CUREB\footnote{\url{https://carleton.ca/researchethics/cureb-b/}, CUREB-B Clearance\#117292}. The study targeted employees who work remotely, fully or partially, during the pandemic.  Participants were asked to complete a questionnaire of $25$ questions divided into four parts, designed to minimize cognitive load and allow participants to fully consider the topics presented \cite{b24}. The first part of the questionnaire collected demographic information, working environment, work experience, IT knowledge, work sector, and the size of the organization's manpower. The second part focused on organizations' preparedness to shift to the WFA scheme and their procedures to defend against detected attacks and increase employee security awareness. The third part aimed to understand common security-related behaviors of employees while working remotely. The final part sought to determine employees' awareness of common cyberattacks and vulnerabilities. 
We follow the recommended three-stage process by Dillman \cite{b25} to pre-test the questionnaire.  First, the authors' team discussed the clarity of the questionnaire and the motivation to identify unclear questions, ambiguous instructions, or other issues before public dissemination. Colleagues and experts in the field then reviewed the questionnaire to uncover potential misunderstandings or unexpected outcomes. Finally, we performed pilot tests with eight graduate-level colleagues to identify any flaws in the questionnaire and determine its appropriate length. The user study was finalized after incorporating feedback from the previous stages. We used Qualtrics \cite{b23} as an online survey tool for this study.
\vspace{-.5cm}
\subsection{Participant Recruitment and Data Quality Assurance}
Recruiting participants can be challenging in this type of research. We advertised the questionnaire on various social media platforms to attract a diverse range of participants from different types of organizations. This was done through Carleton University accounts and our accounts.
To ensure the quality of the data, we took multiple precautions. Since our participants came from different fields, with varying levels of experience and IT knowledge, we ensured that the questionnaire was written in simple, plain English to reach as many participants as possible. Before starting the questionnaire, we briefly explained the study's goal and the areas participants should expect to be asked about. 
To ensure that participants were answering questions genuinely and not just filling out answers arbitrarily, we included challenging questions in each section of the questionnaire. For example, we included a question asking participants to choose the word "Bule" from the answer options if they were reading it. Participants who reported working in person at their offices or failed to answer the challenging questions were excluded from the analysis.
\vspace{-.5cm}
\subsection{Participant Demographics}

This study aims to investigate employees with varying levels of IT knowledge in different fields who worked remotely from any location during the COVID-19 pandemic. The online survey study was conducted between March and June $2022$, with $46$ participants, and had an average completion time of approximately $8.7$ minutes. 
It is worth noting that the vast majority of participants, around $89.13\%$, were working and residing in Canada.

\subsubsection{\textbf{Organization types distribution}} 
\begin{figure}[!ht]
\vspace{-.6cm}
    \includegraphics[width=9 cm, height=2.8cm]{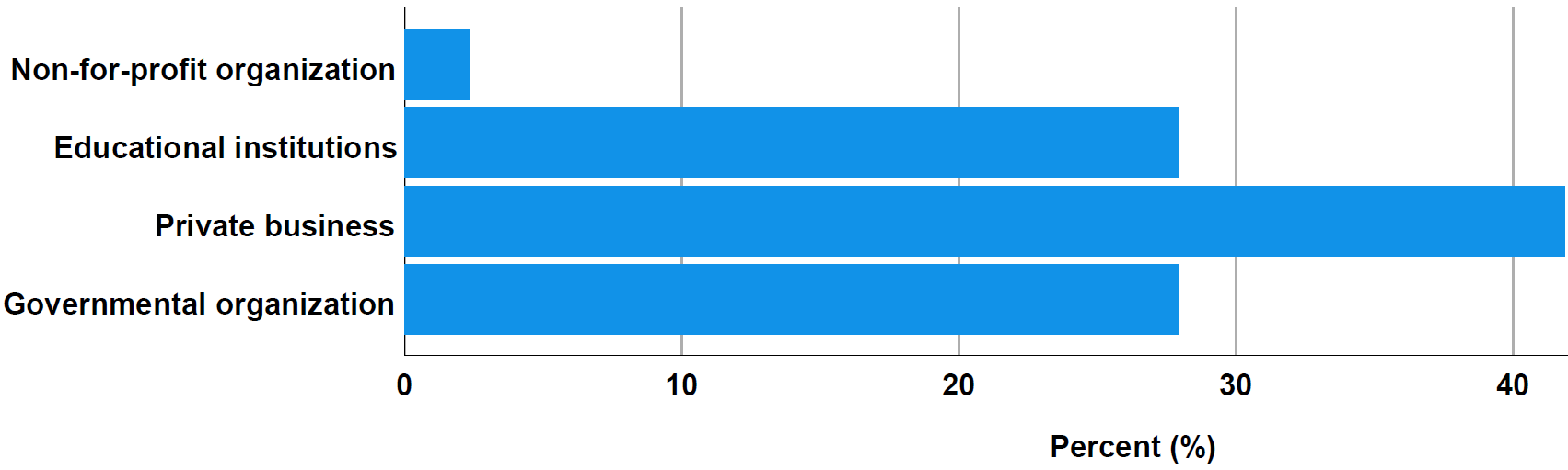}
    \vspace{-.8cm}
    \caption{Distribution of participants' percentage per organization type.}
    \label{fig:OrgType}
\end{figure}

People participating in this study are from different organization types, as shown in Figure \ref{fig:OrgType}. The largest percentage of participants, $42.22\%$, worked in private businesses, followed by 28.89\% in governmental organizations and $26.67\%$ in academic-educational institutions. Only one participant worked for a non-profit organization. This diverse representation of organizational types would provide a better understanding of whether the policies implemented by these organizations have an impact on their readiness and security.

\subsubsection{\textbf{IT knowledge distribution}}  
\begin{figure}[!ht]
    \includegraphics[width=8 cm, height=2.8cm]{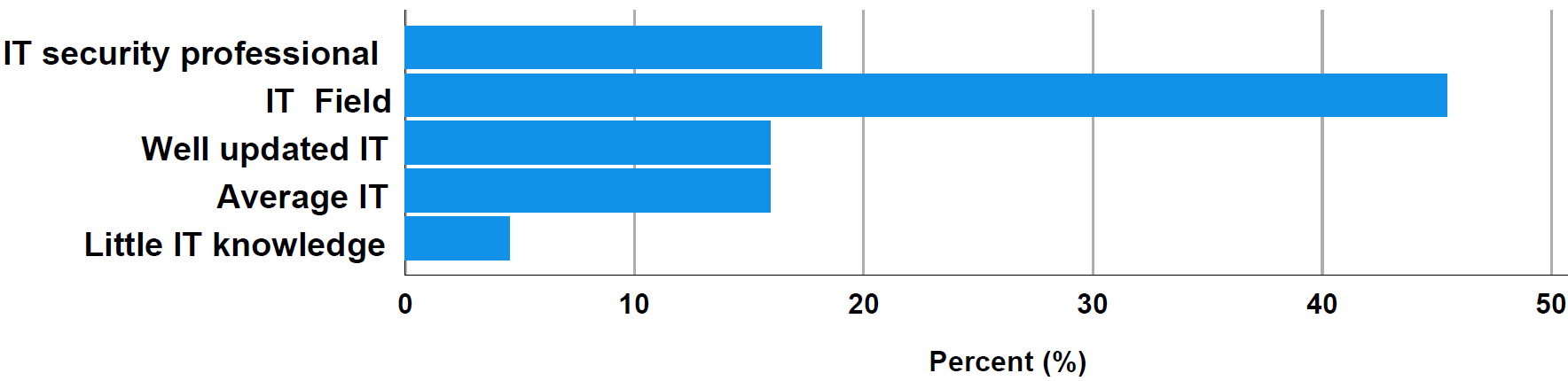}
    \vspace{-.4cm}
    \caption{Distribution of participants' IT knowledge levels.}
    \label{fig:ItKnowledge}
\end{figure}

The participants in this study had varying levels of IT knowledge, ranging from basic to professional, as shown in Figure \ref{fig:ItKnowledge}. The majority of participants ($65.22\%$) worked in the IT field as programmers, network engineers, or security professionals, while the remaining $34.79\%$ had varying levels of IT knowledge from well-updated to little IT knowledge. 
    \vspace{-.4cm}

\subsubsection{\textbf{Years of experience distribution} }
\begin{figure}[!ht]
    \includegraphics[width=8 cm, height=3cm]{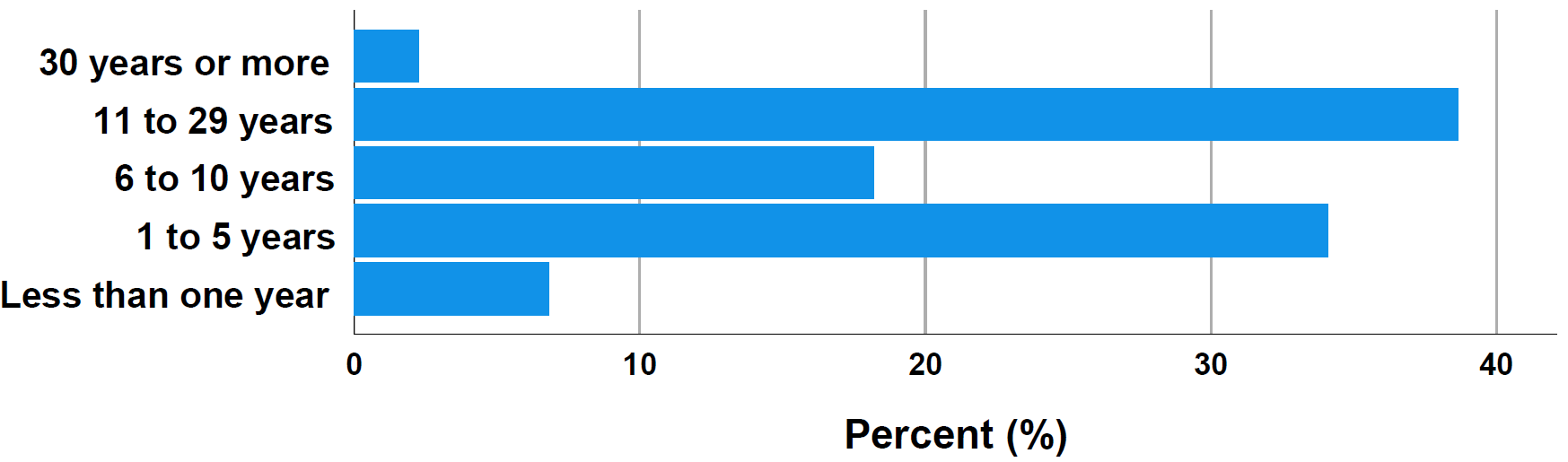}
        \vspace{-.4cm}
    \caption{Distribution of participants' years of experience.}
    \label{fig:YearsOfExperience}
\end{figure}

Figure \ref{fig:YearsOfExperience} displays the distribution of participants' years of experience in the study. The data reveal a wide range of experience levels among the participants. The largest group of participants falls into the category of $11-29$ years of IT experience, closely followed by those with $1-5$ years of experience.
        \vspace{-.4cm}

\subsubsection{\textbf{Employers size distribution}}  
\begin{figure}[!ht]
    \includegraphics[width=8.3 cm, height=3cm]{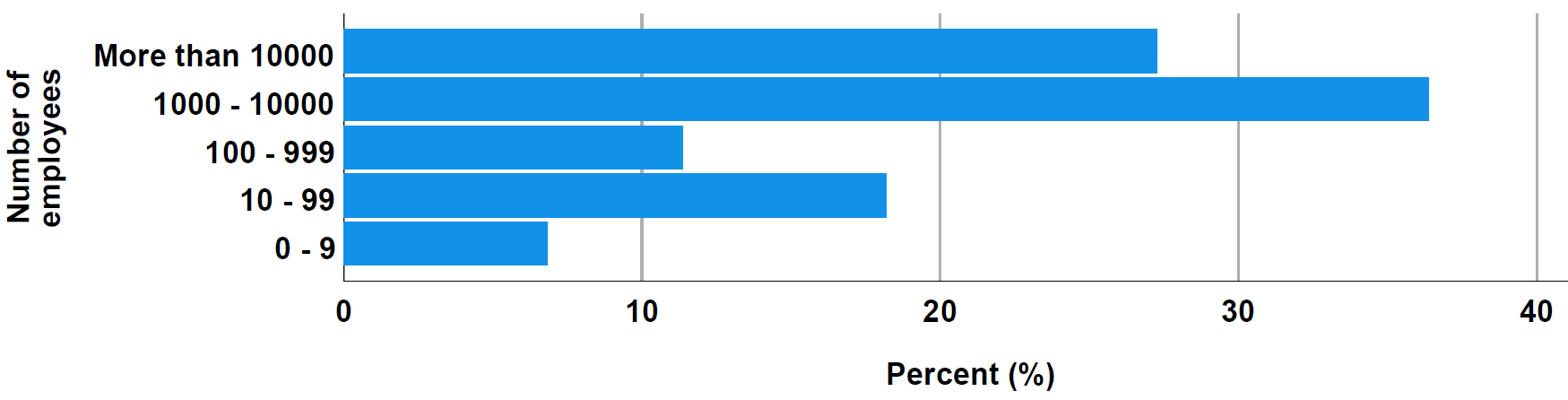}
        \vspace{-.4cm}
    \caption{Distribution of organizations size.}
    \label{fig:EmpStrength}
    \vspace{-.5cm}
\end{figure}

The distribution of employer staff size in terms of the number of employees in the organization is presented in Figure \ref{fig:EmpStrength}. The results indicate that the participants work for organizations of varying sizes, with 37\% of the participants working in large organizations. 
    \vspace{-.4cm}

\subsection{Validity Analysis}
\subsubsection{\textbf{Validity assessment}} To assess convergent validity, a principle component analysis (PCA) \cite{b30a} is conducted. Bartlett's test was significant ($\chi^2 = 507.1 ;~ df=276;~ p~<.001$).  Additionally, the sampling adequacy is evaluated using Kaiser-Meyer-Olkin (KMO), and the obtained value of $.556$ indicated that the collected data were suitable for factor analysis and that the sample size was adequate. Therefore, there are no issues regarding sample size that could affect the validity of the analysis.

\subsubsection{\textbf{Scale reliability of the employees' satisfaction} } 
To evaluate the reliability of the "employees satisfaction" scale, a Cronbach's alpha analysis was conducted. This scale asks employees to rate their employer using a 5-point Likert scale (1: Excellent, 2: Very good, 3: Good, 4: Poor, and 5: I don't know) based on their satisfaction with secure software, IT support, security training awareness, and security policies applied by the employer. The analysis showed that the scale had an \textit{alpha} level of $.90$, indicating an adequate level of inter-item reliability. Furthermore, removing any item from the scale did not improve the \textit{alpha} level, suggesting that each item contributes to the overall reliability of the scale.
\vspace{-.4cm}
\subsection{Theoretical Framework}
The findings of this study align with well-established theoretical frameworks in cybersecurity and human-computer interaction (HCI). Our analysis can be viewed in the context of the following frameworks: 1) The Protection Motivation Theory (PMT) \cite{b33a}, which explains how individuals adopt protective behaviours when confronted with potential threats. This theory provides insight into how employees perceive cybersecurity risks and take protective actions such as avoiding public Wi-Fi. 2)The Technology Acceptance Model (TAM) \cite{b33b}, which explores how users accept and utilize technology. In the context of WFA, this model helps explain the adoption of security tools provided by employers, such as MFA and VPNs.
\vspace{-.4cm}
\section{User-Study Results Analysis}
\label{sec:results_and_analysis}
\subsection{Security-enhancing Tools/Resources}
\subsubsection{\textbf{Status of security tools/resources provided}} 
\label{subsub:ProvidedFacilitiesStacked}
\begin{figure}[!h]
    \includegraphics[width=8.8cm, height=4.8cm]{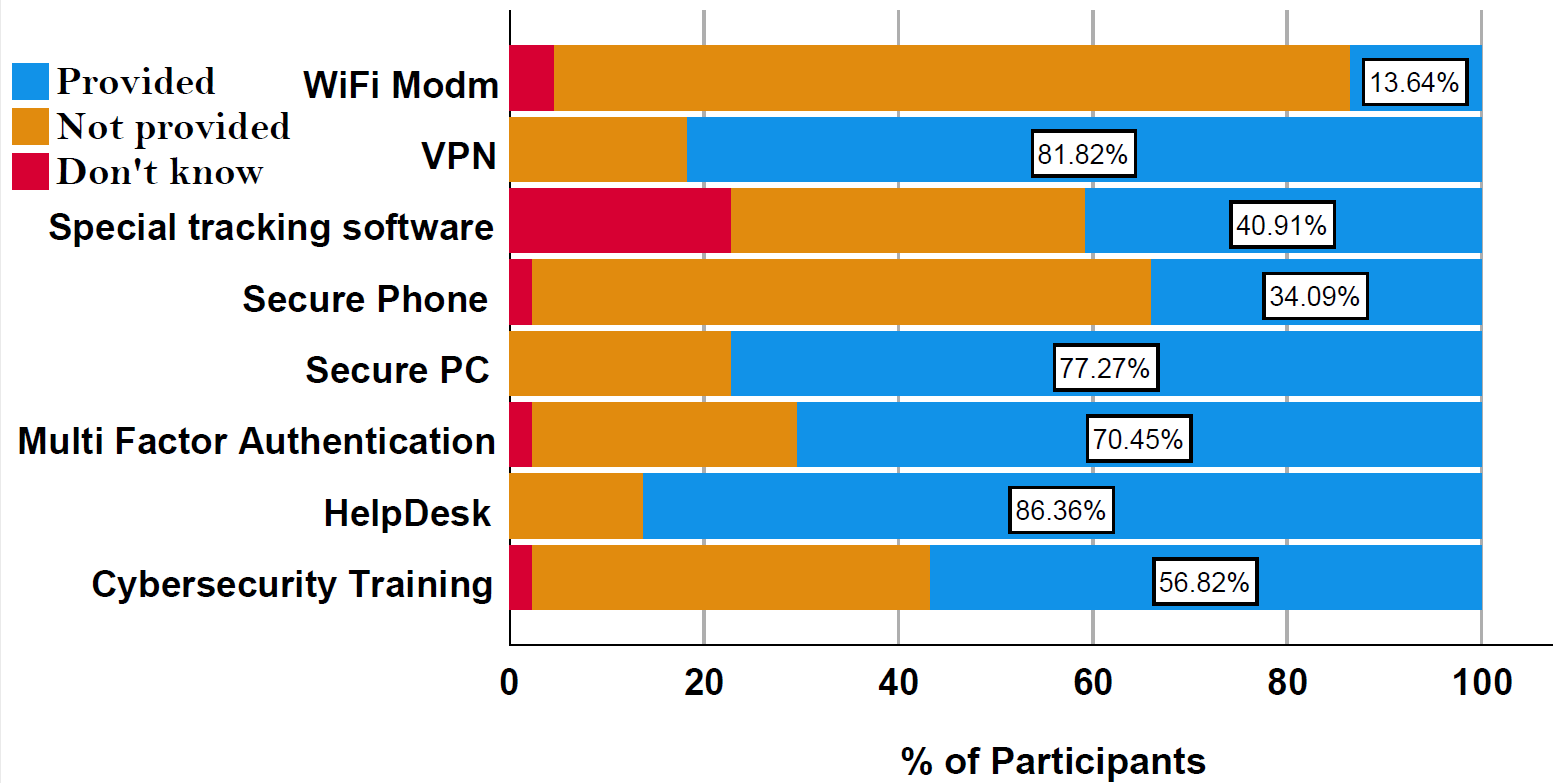}
    \vspace{-.6cm}
    \caption{Tools/resources provided by employers (best viewed in color).}
    \label{fig:ProvidedFacilitiesStacked}
\end{figure}

According to collaborative research \cite{b31a}, a staggering 88\% of data breach incidents stem from errors made by employees. 
To mitigate this risk, security-enhancing tools and awareness training can be provided to help employees identify cyberattacks, practice good cyber hygiene, and comprehend the security risks associated with their behavior. In our study, we asked participants about the tools and resources provided by their employers, and their responses are shown in Figure \ref{fig:ProvidedFacilitiesStacked}. 
The figure displays the percentage of participants who answered whether  a particular tool or resource is provided. It is concerning that approximately $40\%$ of the participants reported not receiving any security training, despite the fact that many security organizations and professionals recommend it to remote workers \cite{b34}, \cite{b36}. However, organizations that prioritize an excellent organizational culture and receive support from top management tend to focus on controlling user access to corporate networks using VPN, multi-factor authentication (MFA), and secure PCs, and providing a help desk.

\subsubsection{\textbf{Status of resource/tools provision across organizations}}
The Chi-square test  is performed to determine if there is a significant difference between the responses of provided and not-provided tools/resources across organizations. Table \ref{tab:facilityProvidedChiSquare} presents the Chi-square values and their corresponding p-values. The results indicate that WiFi modems followed by secure phones had the highest Chi-square values as tools/resources not provided by employers, while the help desk followed by security training awareness had the highest Chi-square values as tools/resources provided by employers. However, the special tracking software had a \textit{p-value} of $.307$, which is greater than $.05$ and suggests that there is no significant difference between the provided and not-provided responses across organizations for this tool/resource.
\vspace{-.4cm}
\begin{table}[!h]
    \centering
    \caption{Chi-square test showing the difference significance between provided and non-provided tools/resources across organizations}
    \begin{tabular}{|l|l|l|l|l|}
    \hline \hline
    Facility & $\chi^2$ & P-value & Status  \\ \hline
    WiFi Modem          &     47.091       &  $ \textless .001$   & Not-provided     \\ \hline
    Secure Phone          &    24.864        &  $ \textless .001$    & Not provided    \\ \hline
    MFA        &     31.41       &   $ \textless .001$     & Provided  \\ \hline
    Help Desk          &     23.273       &   $ \textless .001$    & Provided   \\ \hline
    Cybersecurity Training        &     20.773       &   $ \textless .001$     & Provided  \\ \hline
    VPN               &     17.818       &   $ \textless .001$     &Provided  \\ \hline
    Secure PC             &     13.091       & $ \textless .001$      &Provided    \\ \hline
    Special tracking software      &    2.364         & .307                & -       \\ \hline
    \end{tabular}
    \label{tab:facilityProvidedChiSquare}
\end{table}
    
\subsubsection{\textbf{The relationship between the organization size and tools/resources provided}}
\label{subsub:facilityProvided_vs_SizeChiSquare}
Here, we use demographic information and questions regarding the provided resources/tools to answer the following question: "\textit{How does the organization size (i.e., the number of employees) relate to the IT-specific tool/resource provided?}". A Kruskal-Wallis H (i.e. one-way ANOVA) test reveals that there was a statistically significant difference in the provided cybersecurity training and MFA  between different employer sizes as shown in Table \ref{tab:facilityProvided_vs_SizeChiSquare}. 
Further, the rank biserial correlation (i.e. Spearman test) shows a significant positive bivariate association between the employer size  and providing MFA with a correlation coefficient of $0.518$ and $p<.001$. This suggests that larger organizations are more likely to provide MFA to their employees compared to smaller ones.

\begin{table}[]
    \centering
    \caption{The correlation between resources provided and the organization size. 
    }
    \begin{tabular}{|l|l|l|l|}
    \hline
    Resource/tool & DF & $\chi^2$ & \textit{p-value} \\ \hline
         Cybersecurity training         &  4   &        10.006                                 &      0.04   \\ \hline
              MFA    &  4   &    15.457                                   &    0.04     \\ \hline
    \end{tabular}
    \label{tab:facilityProvided_vs_SizeChiSquare}
    \vspace{-.5cm}
\end{table}
\vspace{-.4cm}
\subsection{Employees Satisfaction and Behavior} 
\subsubsection{\textbf{Assessing employee satisfaction with employer's cybersecurity support and IT resources}}
\label{subsub:SatisfyWithSupport}
\begin{figure}[!h]
    \includegraphics[width=8.5 cm, height=2.7cm]{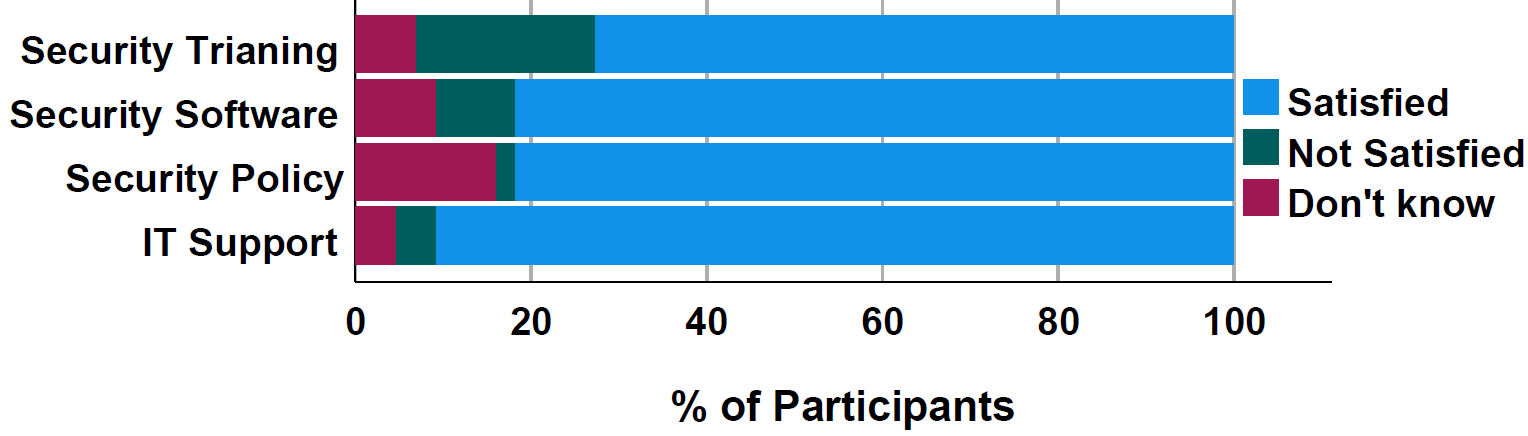}
    \vspace{-.3cm}
    \caption{Employees' satisfaction with security support provided by employers.}
    \label{fig:EmpRedayRateStacked}
\end{figure}

Participants were asked about their level of satisfaction with their employer's support in terms of security training, software, security-specific policies, and IT support. The majority of participants reported feeling satisfied with the security support provided by their employers, as shown in Figure \ref{fig:EmpRedayRateStacked}. We also sought to determine whether there was a correlation between IT-related resources provided by employers and employee satisfaction levels. Our analysis revealed a significant positive correlation between the IT help desk provided by employers and employee satisfaction with IT support. Furthermore, the Spearman test found a strong positive correlation between employee satisfaction with cybersecurity training and the amount of cybersecurity training provided by their employers, with a correlation coefficient of .590 and a $p-value$ of less than $.001$.
        \vspace{-.4cm}
    

\subsubsection{\textbf{Participants’ willingness to contact the IT team for support}}
\label{subsub:YourReaction_ContactIT}
\begin{figure}[!h]
    \centering
    \includegraphics[width=8.3cm, height=2.7cm]{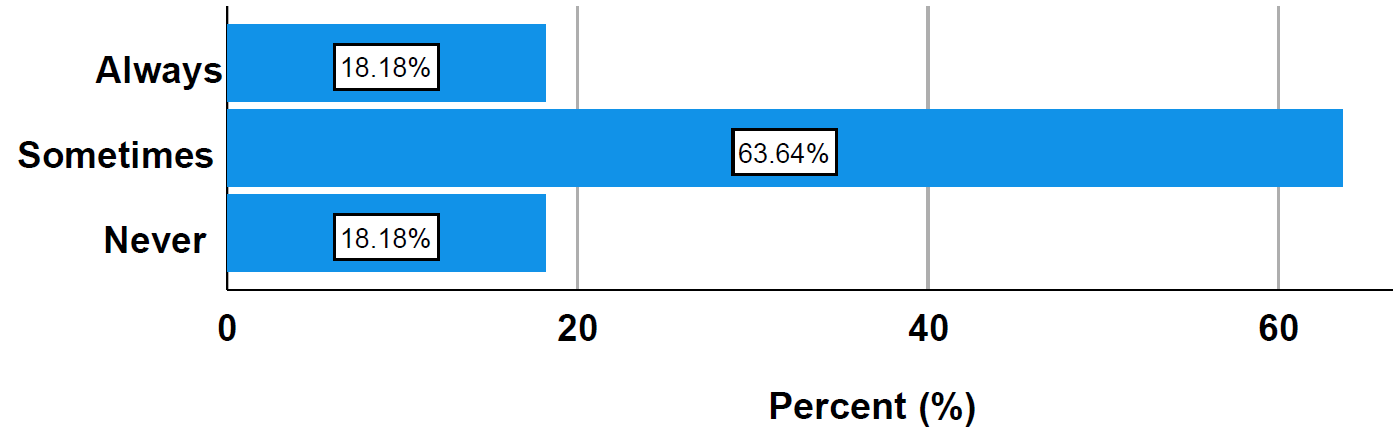}
    \vspace{-.4cm}
    \caption{Participants' behavior in contacting IT team for technical issue}
    \label{fig:YourReaction_ContactIT}
\end{figure}

Participants were asked whether they would contact their IT team for support when needed. Figure \ref{fig:YourReaction_ContactIT} summarizes their responses. As shown in the figure, only $18\%$ of the participants reported that they would always contact the IT team for technical assistance. The majority of the respondents reported that they would sometimes contact the IT department, while $18\%$ reported that they would never ask for IT support from their organization.
    \vspace{-.4cm}


\subsubsection{\textbf{The relationship between employee satisfaction and IT support requests}}
\label{subsub:EmpITSupportSatify_vs_ContactIT}
\begin{figure}[!h]
    \centering
    \includegraphics[width=8.5cm, height=2.7cm]{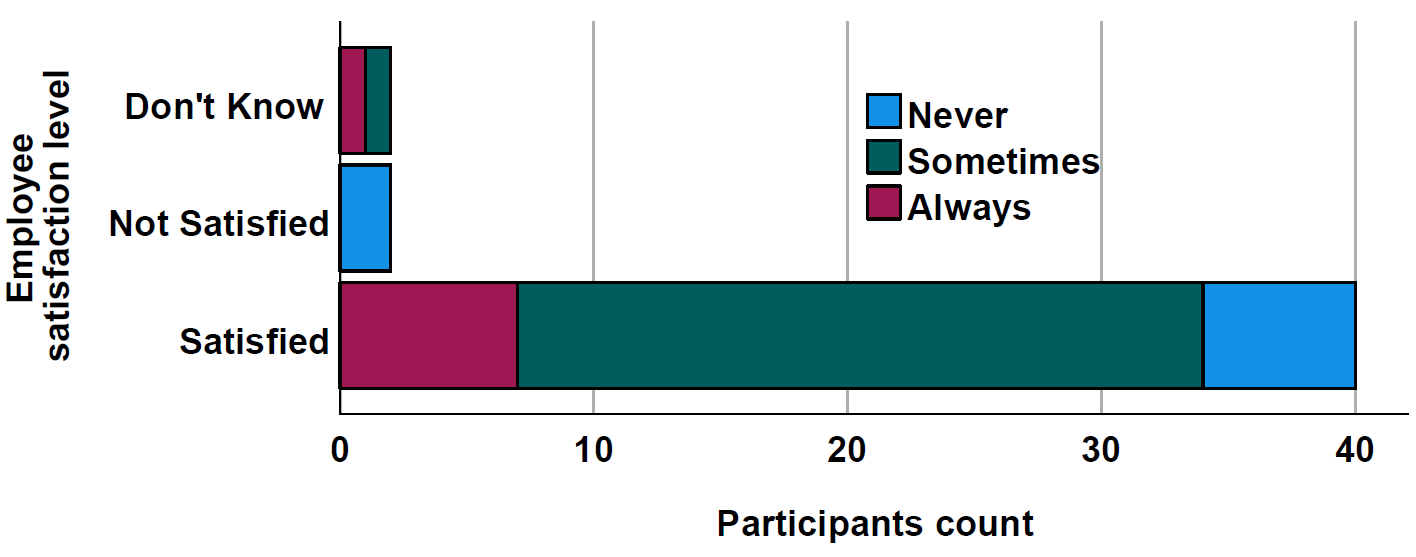}
    \vspace{-.4cm}
    \caption{Relation between employee satisfaction with IT support and tendency to contact IT team for assistance.}
    \label{fig:EmpITSupportSatify_vs_ContactIT}
\end{figure}

From Figure \ref{fig:EmpRedayRateStacked}, we can see that most participants expressed a high level of satisfaction with their employer's IT support. However, the question arises: \textit{ Does employee satisfaction correlate with their tendency to contact the IT team for support?} Figure \ref{fig:EmpITSupportSatify_vs_ContactIT} provides insight into this question. We found that although the majority of employees are satisfied with their employer's IT support, only $17\%$ of them would contact the IT team when faced with technical problems, while $15\%$ would never reach out for support. Instead, the majority of satisfied employees ($67\%$) would only occasionally contact the IT support team. All unsatisfied employees would never ask for help. However, Figure \ref{fig:EmpITSupportSatify_vs_ContactIT} also reveals that $7\%$ of participants who are satisfied with their employer's IT support would never contact the IT team for help. When we asked these participants why they chose not to contact IT, their responses varied. Some participants reported that they could fix problems themselves, while others said that they prefer to ask their colleagues or friends for help. An interesting response was from a participant who said, "Most of the time, I know better than the IT Support Team."
        \vspace{-.4cm}
   


\subsubsection{\textbf{Employees' security behaviors and their impact on organizational security}}
\label{subsub:EmpBehaviours}
\begin{figure}[!h]
    \vspace{-.8cm}
    \includegraphics[width=9.3 cm, height=5cm]{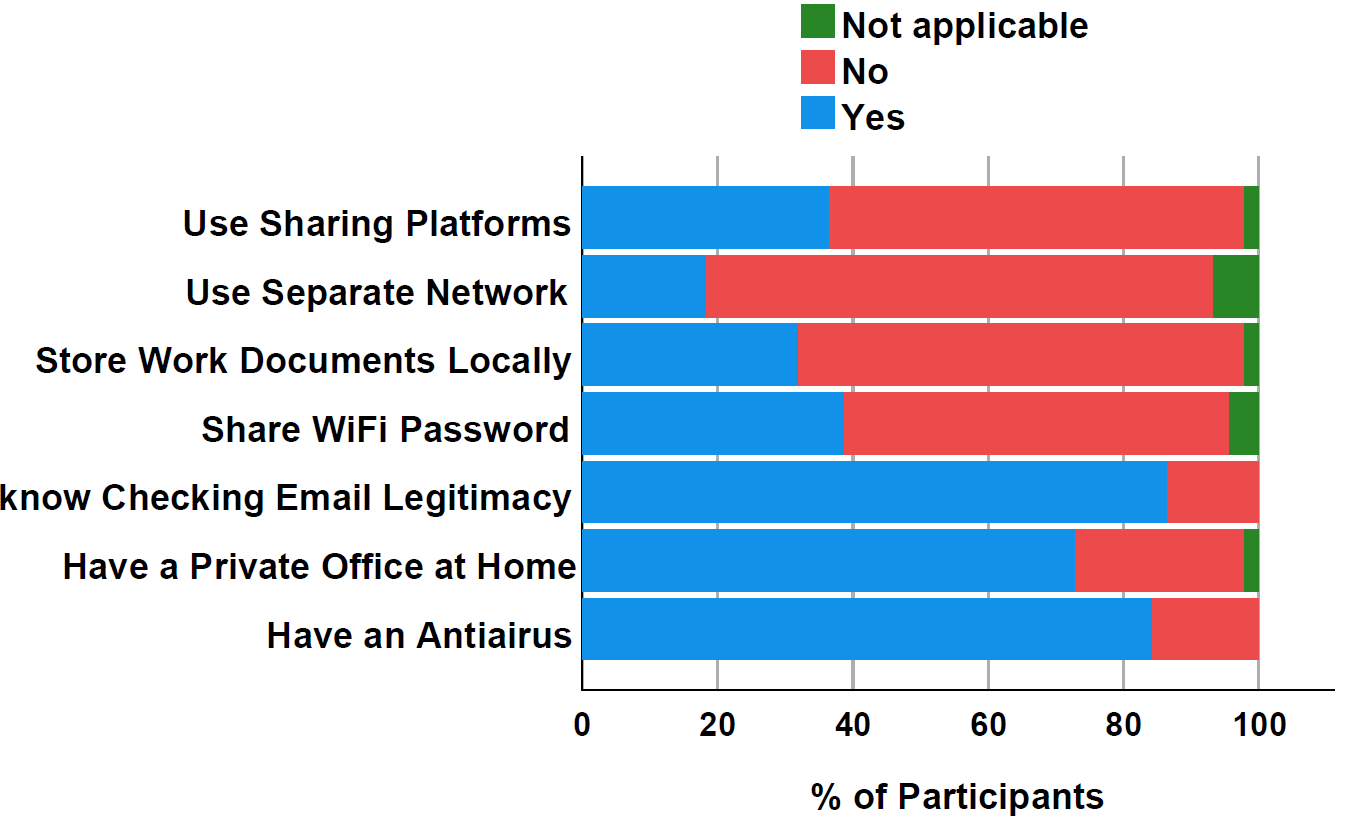}
    \vspace{-.7cm}
    \caption{Comparison of employee behavior across different practices.}
    \label{fig:EmpBehaviours}
\end{figure}

Ensuring the safety of an organization's network and system is a shared responsibility between employees and the IT team. To gauge employees' security awareness and how their behavior may affect the security of the organization, we asked participants about their security-related practices. The questions included their ability to identify legitimate emails, whether they used a separate network for work, whether they used anti-virus software, whether they worked in a private office, whether they shared WiFi passwords with others, and whether they stored work documents on their local computers. Figure \ref{fig:EmpBehaviours} presents the percentages of employee behavior in different practices.
The results indicate that the majority of employees do not practice using a separate network for work, even though it would help mitigate threats. 


\subsubsection{\textbf{The relationship between employees' security attitude and unusual system behavior}}
\label{subsub:EfctEmpBehv_SingsNoticed}
To measure the impact of employee security  attitude on the number of unusual behaviors of their systems, we conducted a multi-way ANOVA and calculated the partial Eta squared ($\eta^2$) for behaviors such as sharing WiFi passwords, using public WiFi, and using public sharing platforms. The result showed that $\eta^2=.132$  and $p=.023$, indicating that $13.2\%$ of the variation in the number of unusual system behaviors noticed by participants is explained by the variation of the mentioned employer behaviors.
   

\subsubsection{\textbf{The relationship between employees' satisfaction with security training and ability to verify email legitimacy}} 
\begin{figure}[!h]
    \centering
    \includegraphics[width=6.2 cm, height=3.2cm] {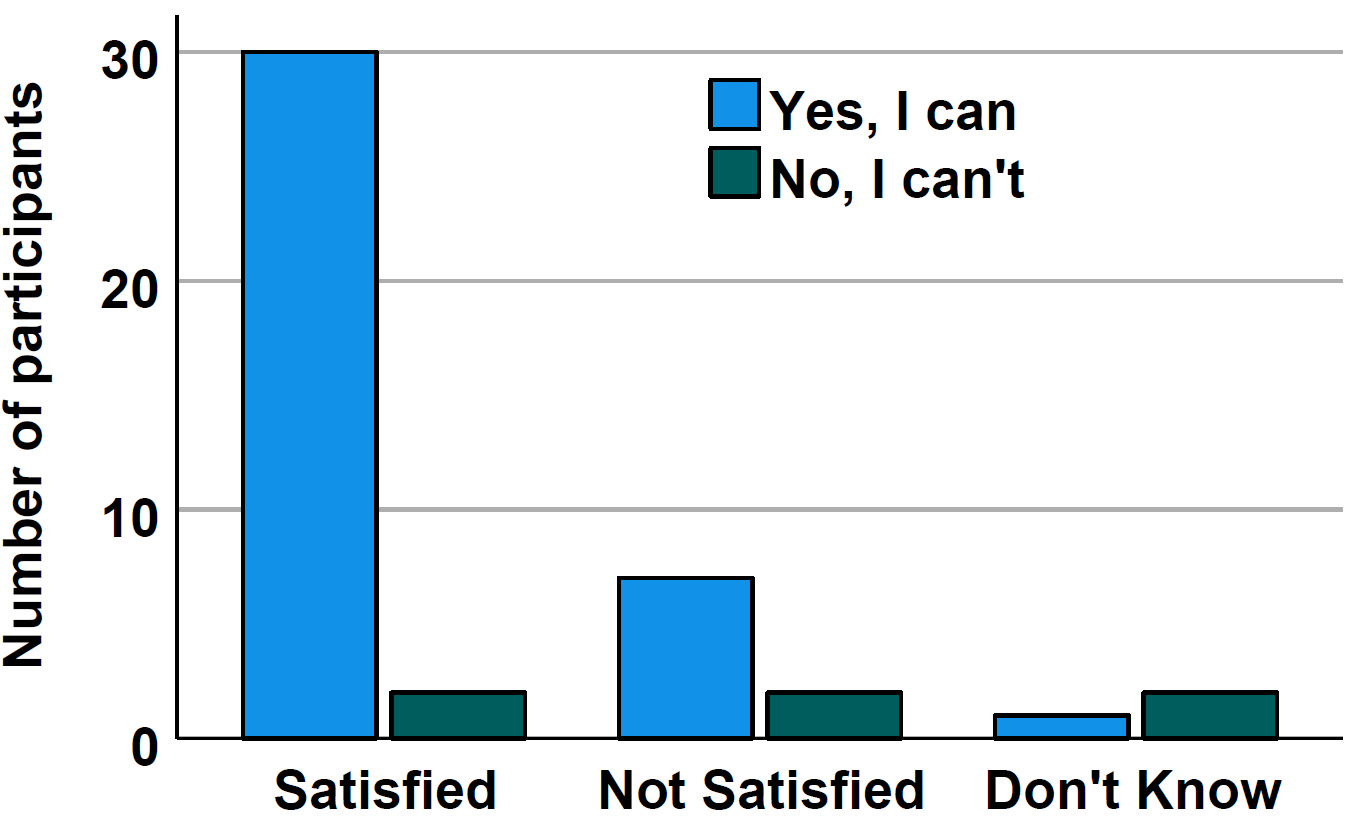}
    \vspace{-.2cm}
    \caption{Relationship between employees' ability to check email legitimacy and their satisfaction with organizational security training.}       \label{fig:SecTrainSatisfy_vs_CheckEmailsLegitimacy}
\end{figure}

A significant relationship was found between employees' satisfaction with their employer-provided security training and their ability to verify the legitimacy of emails, as shown by the Spearman correlation analysis. The correlation coefficient was found to be $.341$ and the \textit{p-value} was $.024$, indicating a strong correlation. This finding highlights the importance of providing security awareness training to employees.
Figure \ref{fig:SecTrainSatisfy_vs_CheckEmailsLegitimacy} provides a summary of the correlation between these two variables. This observation seems intuitive, as employees who receive adequate security training are more likely to be able to identify and respond to suspicious emails effectively. It also emphasizes the need for organizations to invest in comprehensive security awareness training programs to mitigate the risks associated with cyberattacks.
        \vspace{-.4cm}

\subsubsection{\textbf{Employees' perception of public WiFi security}}
\label{subsub:EmpPercpWifi}
The study found that a majority of participants, $57.56\%$, perceive public WiFi to be unsafe for work without using a VPN. Furthermore, $20\%$ of the participants report that they never use public WiFi, suggesting that they take security seriously. In general, the results indicate that employees are aware of the potential security risks associated with public WiFi networks.


\subsubsection{\textbf{Employees' experience with hacking and system modifications}}
\label{subsub:EmpExpWithHacking}

According to the study results, a majority of participants (66\%) reported that they had not experienced any hacking incidents in the previous year. However, a significant percentage (73.84\%) noticed modifications to their systems, which could indicate a potential hack. Note that even though these modifications may not necessarily be the result of a hack, they could still pose a security risk to the organization. Therefore, it is essential to encourage employees to report any suspicious activity to the IT team to prevent any potential security breaches.

\subsection{Employers Policies and Behavior}

\subsubsection{\textbf{Status of employer's policies}}
\label{subsub:EmpPoliciesStatus}
We asked participants whether their employer’s policies allow the use of public sharing platforms such as Google Drive, Dropbox, etc., and public WiFi networks while working remotely. Table \ref{tab:EmpPoliciesAllow} summarizes the results. The majority of participants (63.6\%) reported that the use of public sharing platforms is not allowed, while 25\% of participants reported that the use of public WiFi networks is not allowed. Interestingly, 25\% of the participants reported that they do not know if their employer allows the use of public WiFi networks.
\vspace{-.5cm}
\begin{table}[!h]
    \centering
    \caption{Employee policies allowing the use of public platforms and WiFi}
    \label{tab:EmpPoliciesAllow}
    \vspace{-.3cm}
    \includegraphics[width=8cm, height=2.8cm]{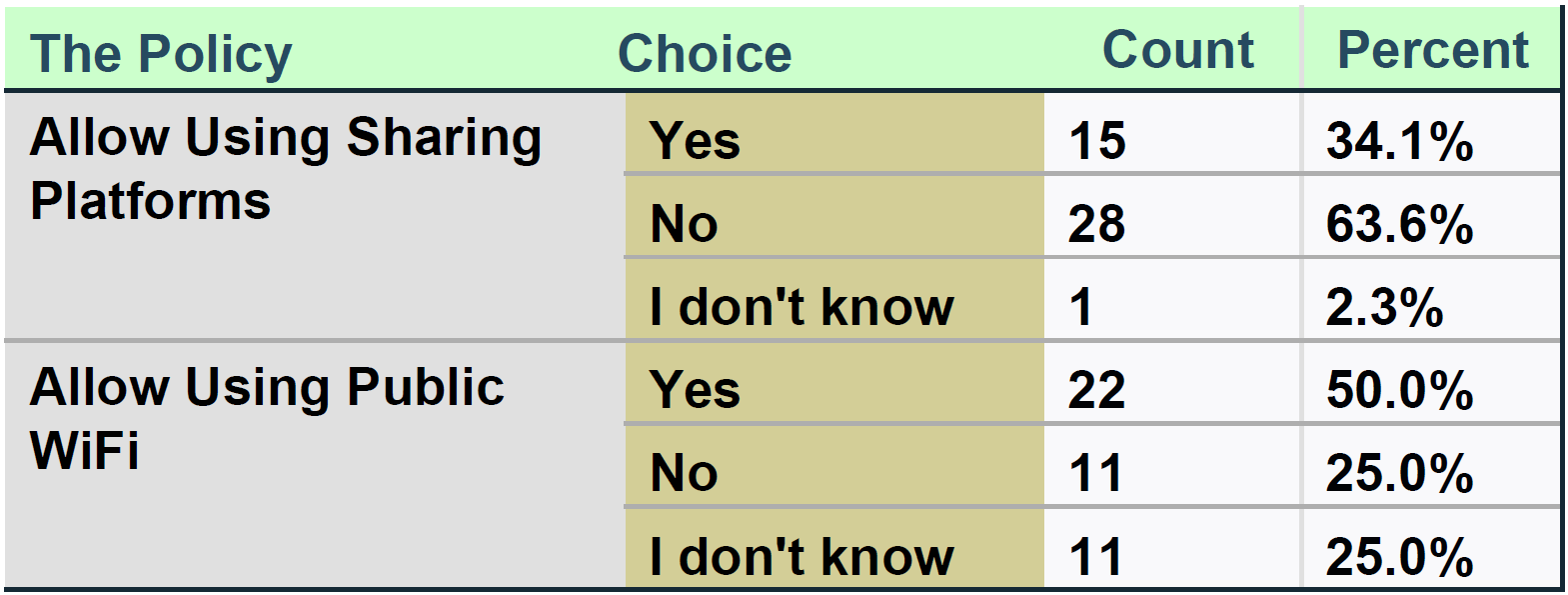}
\end{table}

\subsubsection{\textbf{Compliance with employer's policies and instructions}}
\label{subsub:EmpComplianceWithPolicies}
\begin{figure}[!h]
    \centering
    \includegraphics[width=8.4 cm, height=2.4cm] {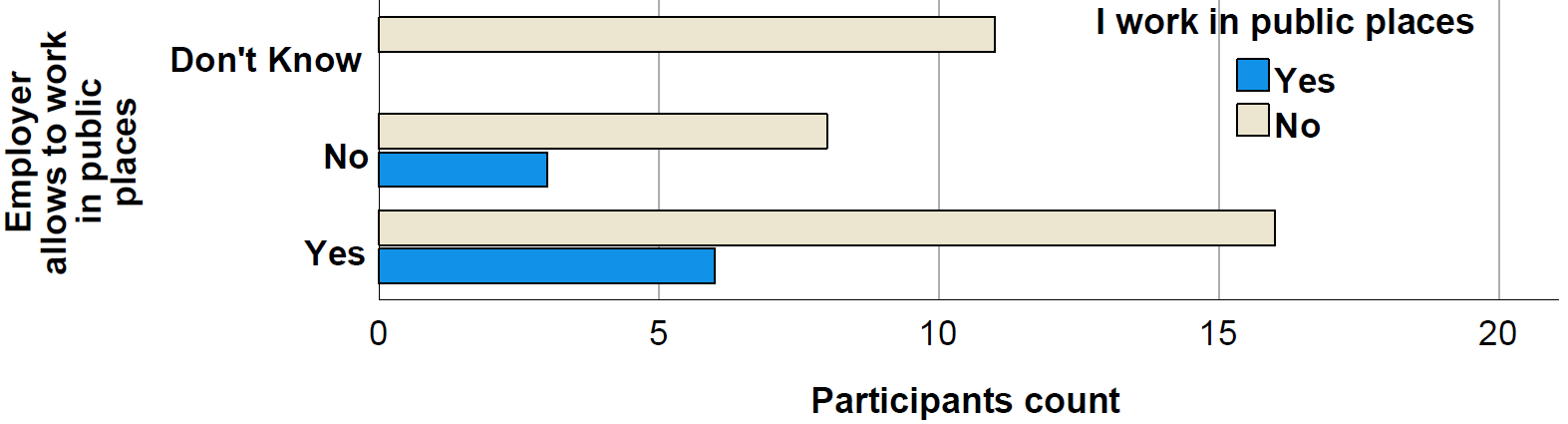}
    \vspace{-.4cm}
    \caption{Employer stance on public WiFi and employee usage.}       \label{fig:EmpAllowsPublicPlaces_vs_YouUsePublicPlaces}
\end{figure}

Our findings show that employees generally comply with their employer’s policies and instructions regarding the use of public sharing platforms and public places (i.e.WiFi) networks. 
The use of \textbf{sharing platforms} by employees and their compliance with employer policies have been examined through a Spearman correlation analysis. The results revealed a significant relationship between the employer's policy on sharing platform usage and employee compliance. With a correlation coefficient of $18.8$ and a \textit{p-value} of less than $.01$, the findings indicate a strong level of compliance among employees
In order to evaluate employee compliance with regards to \textbf{WiFi usage}, Figure \ref{fig:EmpAllowsPublicPlaces_vs_YouUsePublicPlaces} illustrates the stance of employers on the use of public WiFi, as well as the responses provided by participants regarding their personal usage. In particular, the figure reveals that a significant percentage of participants refrain from using public WiFi, even when their employers allow its use. Interestingly, the data highlights that participants tend to abstain from using public WiFi when they "Don't Know" their employer's stance on the matter.
    \vspace{-.3cm}

\subsubsection{\textbf{The relationship between IT knowledge/experience and compliance with employer policies}}
\label{subsub:ITKnldg_vs_PoliciesComplience}
It is noteworthy that we did not find a significant relationship between employees' IT knowledge/years of experience and compliance with the employer’s policies and instructions. This finding suggests that, regardless of IT knowledge and experience, employees tend to follow the policies and instructions set by their employers.

\subsubsection{\textbf{Employer's awareness of VPN's importance}}
\label{subsub:EmpAllowPubWiFi_vs_EmpProvideVPN}
\begin{figure}[!h]
    \centering
    \includegraphics[width=8.5 cm, height=2.3cm] {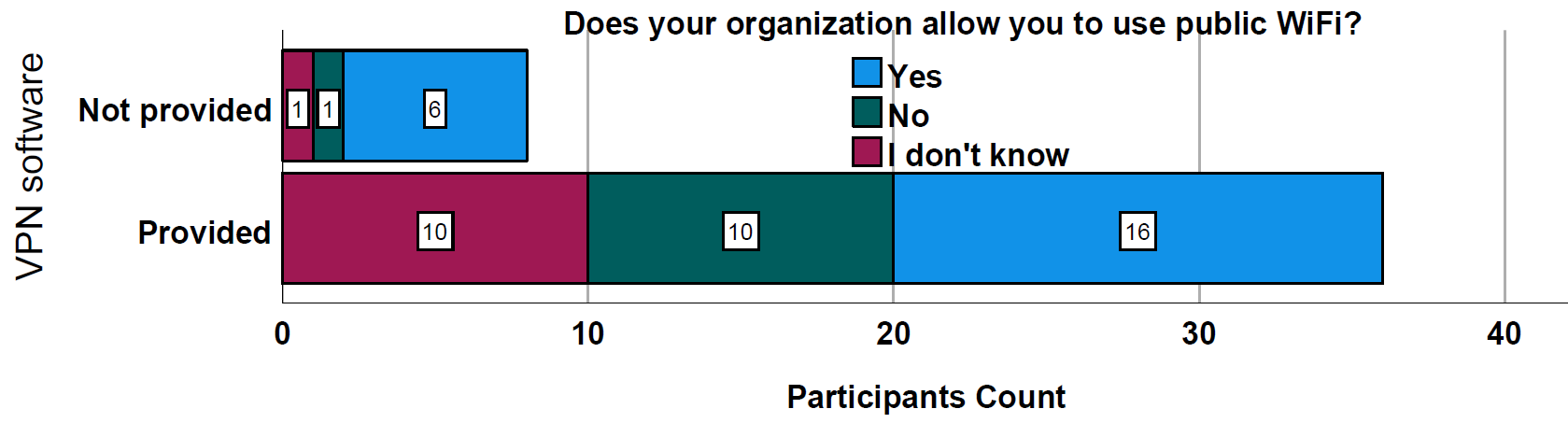}
    \vspace{-.4cm}
    \caption{Public WiFi usage permission and provision of VPN software.}
    \label{fig:EmpAllowPubWiFi_vs_EmpProvideVPN}
    \vspace{-.4cm}
\end{figure}

Figure \ref{fig:EmpAllowPubWiFi_vs_EmpProvideVPN} shows the percentage of participants whose organizations allowed them to use public WiFi and provided VPN software. Despite the fact that a high percentage of participants were allowed to use public WiFi, they used VPNs provided by their organizations to connect to work servers, indicating awareness of the importance of VPNs.

\subsubsection{\textbf{Lack of communication between employers and employees about cyberattacks}}
\label{subsub:OrgType_vs_HaveYouInformedOfAttacks}
\begin{figure}[!h]
    \centering
       \vspace{-.3cm}
       \includegraphics[width=7.8 cm, height=2.8cm]{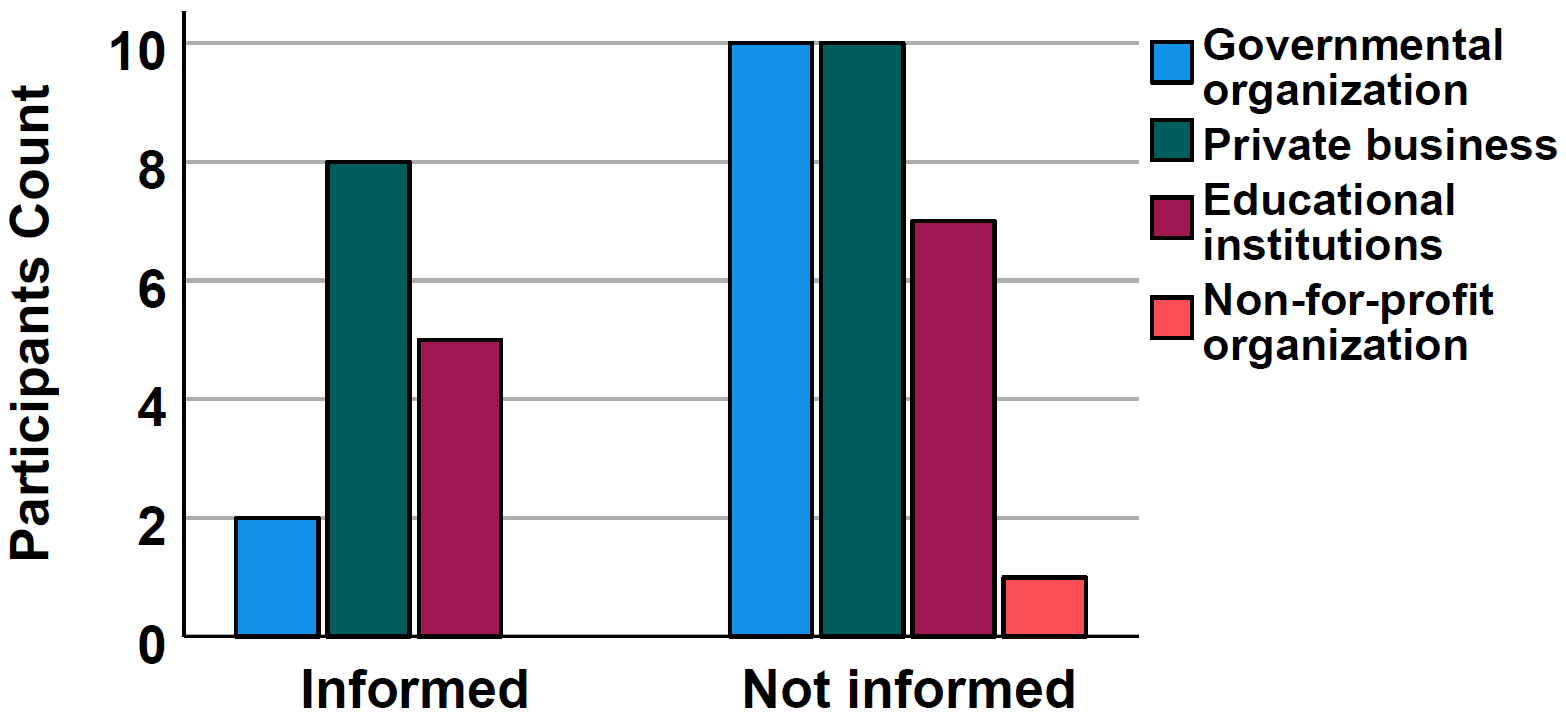}
    \vspace{-.2cm}
    \caption{Participants' response "\textit{Have you been informed of any cyberattack at your organization in the past $12$ months or since the pandemic?}"}
    \label{fig:OrgType_vs_HaveYouInformedOfAttacks}
\end{figure}

The study  found that there is a miscommunication between employers and employees regarding cyberattacks. A descriptive analysis was conducted to determine whether employers generally inform their employees about attacks that occurred during the pandemic. The results show that $79\%$ of the participants from all types of organizations do not receive any notification. Figure \ref{fig:OrgType_vs_HaveYouInformedOfAttacks} illustrates the number of participants who receive information and those who do not. Interestingly, $91.7\%$ of government employees do not get informed about attacks, representing the highest percentage among participants in different organizations.  It is important to note that although there is a possibility that organizations do not inform their employees due to the absence of attacks, it is unlikely that all organizations have remained unaffected by any attacks.
\subsubsection{\textbf{Relationship between employees' IT knowledge and being informed about attacks}}
\label{subsub:EmpITKnldg_vs_AttackInformed}
The results of the Spearman rank test revealed a negative correlation between IT knowledge and being informed about attacks, with a coefficient of $-.313$ and a significance level of $p=.038$. In other words, employees with higher IT knowledge are less likely to be informed about attacks than those with lower IT knowledge. These findings highlight the need for better communication and education among employers and employees regarding cyberattacks, as even those with a higher level of IT knowledge may not be fully informed about the security risks facing their organization. 



\subsubsection{\textbf{Organizational defense procedures in response to cyberattacks}}
\label{subsub:EmpRespAttck_Stacked}
\begin{figure}[!h]
    \includegraphics[width=8.6cm, height=2.5cm]{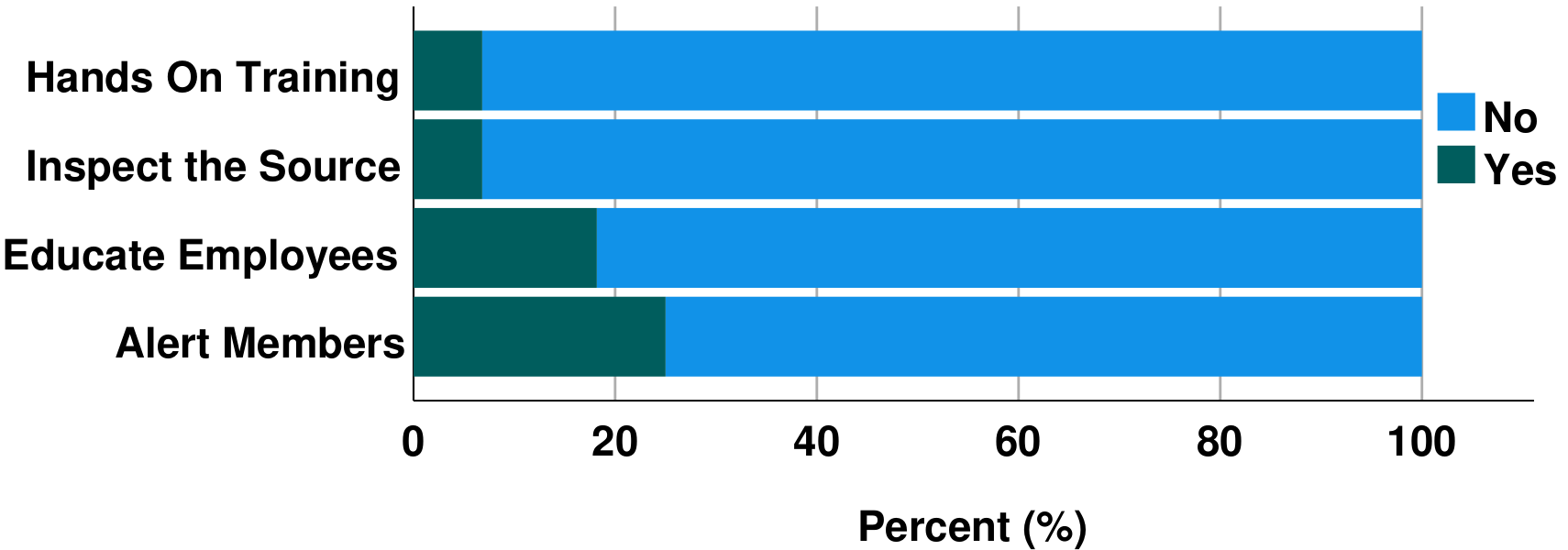}
    \vspace{-.4cm}
    \caption{Organizational defense procedures in response to cyberattacks.}
    \label{fig:EmpRespAttck_Stacked}
\end{figure}

The increase in  cyberattacks during the pandemic  is due in large part to the decision and behavior of organizations during the pandemic \cite{b33}. Participants were asked about the defense procedures of their employers in response to threats they encountered without revealing those threats. Figure \ref{fig:EmpRespAttck_Stacked} shows the procedures and their percentages. All the procedures we specified in the questionnaire are seldom implemented across all organizations, as can be seen from the figure. Although participants were given the opportunity to mention procedures not listed in the options, they have not recalled other actions their organizations may apply in response to attacks. We note that there may be some defense procedures that were implemented by organizations that the respondents may not be aware of. Recently, the Canadian Centre for Cybersecurity in \cite{b37} highly recommended developing an incident response plan to help mitigate the risk of cybersecurity incidents. 
    \vspace{-.4cm}
    \vspace{-.4cm}
\section{User-Study Findings Summary and Limitations}
\label{sec:limitations}
\subsection{User-Study Findings Summary}
This study reveals several key findings, which can be summarized as follows:
\begin{itemize}
    \item Despite the importance of cybersecurity awareness and education, a high percentage of participants did not receive security training. Instead, organizations prioritized user access control, VPN, MFA, and providing a help desk.
\item Large organizations were more likely to provide MFA to their employees compared to small ones.
\item Employees were generally satisfied with the security support and IT resources provided by their employers. There was a strong positive correlation between employee satisfaction with cybersecurity training and the amount of training provided by their employers. Additionally, there was a significant positive correlation between the IT help desk provided by employers and employees' satisfaction with IT support.
\item Only a small percentage of employees reported that they would always contact the IT team for technical assistance when needed, while the majority reported that they would only occasionally contact the IT department or never ask for IT support from their organization.
\item Participants who received security training were more likely to identify and respond to suspicious emails effectively.
\item Participants generally complied with their employer's policies and instructions regardless of their IT knowledge and experience.
\end{itemize}
\vspace{-.5cm}
\subsection{User-Study Challenges and Limitations}
The following limitations should be taken into account when considering the findings or recommendations mentioned in this paper. 
\subsubsection{\textbf{participants recruitment}}
The user study was designed with the utmost consideration for participant anonymity and the protection of their personal data. Therefore, we deliberately refrained from asking participants for information like email addresses, phone numbers, or social media accounts. Additionally, we decided not to offer compensation to avoid potential privacy concerns. At first, we were unaware of an alternative method for compensating participants without collecting their contact information, which ultimately affected the number of participants who completed the questionnaire. However, later on, we discovered that online participant recruitment tools such as Prolific \cite{b30} provide a solution for this purpose.  Employing such tools at the beginning of the study would have resulted in an increased number of participants for the study.

\subsubsection{\textbf{Time constraints and questions:}}
In today's fast-paced world, time is a valuable resource for individuals. To respect the time constraints of the participants, the questionnaires must be concise, taking no more than $5$ to $15$ minutes to complete. Consequently, the limited time available influenced the number of questions we could ask participants. Ideally, a longer questionnaire would have been designed to gather a more comprehensive understanding of organizational readiness, user behavior, and awareness.
\vspace{-.5cm}
\section{Best practice recommendations}
\label{sec:recommenations}
 \subsection{Recommendations Based on User Study Findings}
The recommendations provided in this section are based on the findings derived from our user study. These recommendations are based on the empirical data collected from the participants, offering valuable insights into their behaviors, practices, and perceptions regarding cybersecurity. The recommendations aim to address the specific issues identified in the study, providing actionable steps for organizations to improve their cybersecurity posture. By tailoring these recommendations to the insights gained from the user study, organizations can effectively mitigate risks, improve employee awareness, and establish robust security measures in the context of remote work. 
Although not exhaustive and are unable to guarantee cybersecurity results, these recommendations encompass various areas that can aid in mitigating cybersecurity risks.
\subsubsection{\textbf{Enhancing help desk}} 
Considering the findings in Subsection \ref{subsub:EmpITSupportSatify_vs_ContactIT} related to employees' hesitancy to contact the IT team for support, it is crucial to promote effective communication and encourage engagement between employees and the IT department. To encourage employees to seek IT support when needed, organizations should foster a supportive environment where reaching out to the IT team is encouraged and not seen as a sign of incompetence. Clear and easily accessible channels for IT support should be provided, and the IT team should actively engage with employees, being approachable, friendly, and prompt with their assistance. Additionally, promoting awareness of the IT team's expertise and empowering employees with self-help resources can boost their confidence in seeking IT support, ultimately mitigating potential security risks resulting from unaddressed issues.

\subsubsection{ \textbf{Importance of Antivirus Software}}
According to the study findings, a significant majority of participants, 86\%, reported using antivirus software on their work devices. However, it is concerning that 14\% of the participants were unsure whether antivirus software was installed on their machines. To ensure a robust cybersecurity posture, organizations should prioritize the implementation and regular update of antivirus software on all work devices. In addition, proactive measures should be taken to educate employees about the importance of antivirus protection and provide them with clear instructions on how to check and verify the presence of antivirus software on their devices.

\subsubsection{\textbf{Implementing secure network segmentation for work activities}}
To address the finding presented in Subsection \ref{subsub:EmpBehaviours}, that employees often do not use a separate network for work, organizations should prioritize the implementation of secure network segmentation. This involves the creation of a dedicated network or subnet specifically for work-related activities, separate from the personal networks of employees \cite{b6}. This ensures a higher level of security, protecting sensitive data from potential threats. It can also help to ensure that the corporate network has the resources it needs to operate optimally. This recommendation requires resources and expertise, and it is up to the individual organizations to consider the trade-off between the benefit and the required resources.

\subsubsection{\textbf{Promoting secure behaviors and limit risky practices}}
To address the impact of the security attitude of employees on unusual system behavior highlighted in Subsection \ref{subsub:EfctEmpBehv_SingsNoticed}, organizations should prioritize promoting secure behaviors and limiting risky practices. This can be achieved by emphasizing the importance of not sharing WiFi passwords,  refraining from using public sharing platforms for sensitive information, and using up-to-date anti-virus software. Additionally, encourage the use of secure collaboration tools and implement robust monitoring systems to detect and respond to potential security threats.

\subsubsection{\textbf{Avoiding public WiFi if possible}}  
The study revealed that a significant proportion of participants perceive public WiFi as unsafe for work. Whenever possible, it is recommended to avoid using public WiFi networks \cite{b6a}. However, in the absence of other options, it is highly recommended that remote workers use a VPN to ensure the security of their information \cite{b6a}. 

\subsubsection{\textbf{Using VPN}} 
To address employees' concerns about the security of public WiFi networks as outlined in Subsection \ref{subsub:EmpPercpWifi}, organizations should promote and encourage the use of VPN when accessing work-related resources and sensitive data outside of secure networks. A VPN creates a secure and encrypted connection between an employee's device and the organization's network, protecting data transmission from potential threats on public WiFi networks. By using a VPN, employees can establish a secure tunnel for their internet traffic, ensuring the confidentiality and integrity of their communications.

\subsubsection{\textbf{Using personal hotspots}} 
Using personal hotspots, whether created by a personal phone or a dedicated device, can provide improved security when compared to public WiFi networks \cite{b6b}. These hotspots typically offer encrypted connections, making it more difficult for hackers to intercept data. 

\subsubsection{\textbf{Implementing MFA}} 
Given the significant positive association between organization size and the provision of MFA highlighted in Subsection \ref{subsub:facilityProvided_vs_SizeChiSquare} where larger organizations are likely to implement MFA.  It is recommended that smaller organizations prioritize the implementation of MFA as a security measure. Enabling MFA as an additional step in the login process provides an additional layer of security for accounts. 

\subsubsection{\textbf{Establishing clear and updated security policies}} 
To address the finding, presented in Subsection \ref{subsub:EmpPoliciesStatus}, that a high percentage of participants were unsure about the use of public WiFi networks, organizations should a) develop comprehensive policies that explicitly state whether the use of public WiFi networks and public sharing platforms is allowed or prohibited, b) regularly review and update these policies to reflect evolving cybersecurity threats and organizational needs, c) provide employee education to ensure understanding of the policies, and d) establish clear communication channels to keep employees informed about the policies and any updates. By doing so, organizations can minimize security risks, prevent data breaches, and promote responsible and secure behavior when working remotely.
    
\subsubsection{\textbf{Providing comprehensive security awareness training}} 
Based on the finding in Subsection \ref{subsub:ProvidedFacilitiesStacked}, the employees' satisfaction with security training correlates with their ability to verify the legitimacy of emails. However, it is concerning that approximately $40\%$ of the participants reported not receiving any security training. To address this concern, organizations should implement comprehensive security training programs specifically tailored for remote workers. This training should cover topics such as identifying common cyberattacks, being aware of the latest cybersecurity threats, checking the legitimacy of email, secure home network setup, safe use of public Wi-Fi, and data encryption. By investing in comprehensive security awareness training, organizations can enhance employees' ability to identify and respond to suspicious emails effectively, mitigating the risks associated with cyberattacks and promoting a strong security posture.

\subsubsection{\textbf{Developing an incident response plan}} 
Considering the limited implementation of defense procedures reported by participants, as highlighted in Subsection \ref{subsub:EmpRespAttck_Stacked}, it is crucial for organizations to develop and implement a comprehensive incident response plan. It helps organizations minimize the impact of attacks, reduce downtime, and protect sensitive data and systems. The plan should outline clear procedures for incident detection, reporting, containment, investigation, and recovery. It should also define the roles and responsibilities of the personnel involved in the response process and establish communication channels for effective and timely coordination \cite{b37}. Regular testing and updating of the plan are essential to ensure its effectiveness in addressing evolving threats. By having a well-defined incident response plan in place, organizations can improve their resilience against cyberattacks and minimize potential damage \cite{b37}.

\subsubsection{\textbf{Establishing better communication channels with employees}} 
The study revealed a lack of communication as outlined in Subsection \ref{subsub:OrgType_vs_HaveYouInformedOfAttacks}, with a high percentage of participants not receiving any notification about attacks. This miscommunication poses a risk to organizational security. Employees may not be aware of ongoing cyberattacks, which makes it difficult for them to take appropriate measures to protect sensitive information and systems. To address this issue, organizations are encouraged to establish a systematic process for sharing information with employees to ensure that employees receive timely and relevant information about emerging threats, attack trends, and mitigation strategies. This proactive approach equips employees with the necessary tools and insights to remain vigilant, make informed decisions, and contribute to the overall cybersecurity resilience of the organization.

\subsubsection{\textbf{Encouraging reporting of suspicious activities}} 
To address the finding presented in Subsection \ref{subsub:EmpExpWithHacking}, that a significant percentage of participants noticed modifications to their systems, organizations should a) prioritize encouraging employees to report any suspicious activity to the IT team, b) establish clear reporting channels and provide the option for anonymous reporting to create a culture of transparency and trust, c)  
conduct incident response training to educate employees on recognizing and reporting potential security breaches, and d) ensure prompt response and thorough investigation of reported incidents and provide feedback to employees about the status and outcome of their reports. By encouraging reporting, organizations can proactively identify and address potential security breaches, mitigating risks, and maintaining a secure environment for sensitive information and systems.
\vspace{-.5cm}
\subsection{Findings Alignment with NIST Cybersecurity Framework}
While some of these practices may appear basic, our study reveals that many organizations do not consistently implement the most fundamental cybersecurity measures. We compare these recommendations with established NIST security framework to strengthen these recommendations.

\begin{enumerate}
    \item \textbf{Promote secure behavior through training and awareness programs:} Our study found gaps in security awareness and behaviour among remote workers, with 40\%  of the participants reporting they had not received formal security training. By encouraging a security-conscious culture, organizations can enhance employees’ ability to identify and respond to cybersecurity threats. Additionally, this recommendation is aligned with the NIST Cybersecurity Framework’s Protect function (PR.AT), which stresses the need for training to ensure that personnel know their cybersecurity roles and responsibilities.
    
    \item \textbf{Implement MFA for remote access:} While 70\% of organizations in our study had adopted MFA, a portion of respondents reported it was not consistently enforced. Our best practice recommendation is closely aligned with the NIST Cybersecurity Framework’s Protect function (PR.AC), which advocates using MFA to ensure that only authorized individuals can access organizational systems.
    
    \item \textbf{Use antivirus software and ensure timely updates:} While antivirus software may seem basic, our study found that many participants use it. Antivirus software is a fundamental element of the NIST Cybersecurity Framework’s Protect function (PR.DS), which includes protective technologies such as malware defences to ensure data security.

    \item \textbf{Ensure regular communication about cybersecurity threats and incidents:} Our findings revealed a concerning communication gap, with 79\%  of respondents reporting they were not informed about cyberattacks or security breaches experienced by their organization. This lack of transparency can hinder an organization’s efforts to encourage a proactive security culture. In comparison, the NIST Cybersecurity Framework’s Respond function (RS.CO)  standard both emphasize the importance of clear communication protocols during and after security incidents, ensuring that all relevant stakeholders are informed and prepared to respond to future threats.
\end{enumerate}
    \vspace{-.5cm}
\section{Conclusion}
\label{sec:conclusion}
This study investigates the cybersecurity challenges of the Work-from-Anywhere (WFA) model, using a structured user study to assess both organizational preparedness and employee behaviors in remote work environments. The findings reveal several critical issues, including a need for more security training for remote workers, communication gaps about cyberattacks, and varying levels of compliance with security protocols across different sectors.
Our findings align with the literature on cybersecurity culture, showing how organizational communication and policy dissemination impact employee compliance. The observed communication gaps about cyberattacks highlight the need for a more proactive approach to cybersecurity awareness, particularly in decentralized work environments.
Additionally, our recommendations are aligned with the widely accepted NIST Cybersecurity Framework, ensuring that the proposed practices adhere to recognized standards. By contextualizing our findings within NIST’s guidelines, we provide actionable insights that organizations can implement to improve their cybersecurity posture effectively. This alignment enhances the practical relevance of our study and offers a structured pathway for organizations seeking to strengthen their security policies.
These theoretical and practical contributions offer a solid foundation for future research exploring the intersection of cybersecurity and remote work while providing actionable guidance for organizations navigating the challenges of WFA environments.
Future work can build upon our findings by exploring larger, more diverse samples and analyzing long-term cybersecurity behaviour in remote settings. Further research could also address the privacy paradox, exploring the factors contributing to the discrepancy between employees' concerns about cybersecurity and their actual behaviours.
\vspace{-.3cm}
\section{Acknowledgement}
    This work was supported by the Natural Sciences and Engineering Research Council of Canada (NSERC) through the Discovery Grant program.

\end{document}